\documentclass[conference]{IEEEtran}
\IEEEoverridecommandlockouts
\usepackage{cite}
\usepackage{amsmath,amssymb,amsfonts}
\usepackage{algorithmic}
\usepackage{graphicx}
\usepackage{textcomp}
\usepackage{xcolor}
\usepackage{tikz}
\usepackage{subcaption}
\usepackage{booktabs}
\usetikzlibrary{positioning,shapes.geometric,arrows.meta,calc,fit,backgrounds,decorations.pathmorphing}
\usepackage{orcidlink}
\usepackage[nolist]{acronym}
\usepackage{verbatim}
\usepackage[capitalize]{cleveref}
\usepackage{tikz}
\usepackage{xcolor}
\usepackage[inline]{enumitem}
\usepackage{siunitx}
\usepackage{pifont}
\usepackage{hyperref}
\usepackage{tabularx}

\DeclareSIUnit{\bps}{bps}

\definecolor{naivered}{HTML}{E15759}
\definecolor{alignedblue}{HTML}{4E79A7}
\definecolor{readygreen}{HTML}{59A14F}
\definecolor{paneltint}{HTML}{F5F5F5}
\definecolor{cachegray}{HTML}{E8E8E8}

\Crefname{section}{Section}{Sections}
\crefname{section}{Sec.}{Secs.}
\Crefname{table}{Table}{Tables}
\crefname{table}{Tab.}{Tabs.}
\Crefname{figure}{Figure}{Figures}
\crefname{figure}{Fig.}{Figs.}
\crefname{lstlisting}{List.}{Lists.}

\definecolor{naivered}{HTML}{E15759}
\definecolor{alignedblue}{HTML}{4E79A7}
\definecolor{cachegray}{HTML}{D0D0D0}
\def\BibTeX{{\rm B\kern-.05em{\sc i\kern-.025em b}\kern-.08em
    T\kern-.1667em\lower.7ex\hbox{E}\kern-.125emX}}

\newcommand{\defineacronym}[3]{%
  \expandafter\newcommand\csname #1\endcsname{%
    \ifcsname #1@used\endcsname
      #2%
    \else
      #3 (#2)%
      \expandafter\gdef\csname #1@used\endcsname{}%
    \fi
  }%
}

\begin{acronym}
    \acro{MoQ}{Media over QUIC}
    \acro{MOQT}{Media Over QUIC Transport}
    \acro{ABR}{Adaptive Bitrate}
    \acro{HEVC}{High Efficiency Video Coding}
    \acro{CMAF}{Common Media Application Format}
    \acro{MSF}{MOQT Streaming Format}
    \acro{CMSF}{CMAF compliant MOQT Streaming Format}
    \acro{HAS}{HTTP Adaptive Streaming}
    \acro{DASH}{Dynamic Adaptive Streaming over HTTP}
    \acro{HLS}{HTTP Live Streaming}
    \acro{CDN}{Content Delivery Network}
    \acro{PTS}{Presentation Timestamp}
    \acro{MSE}{Media Source Extensions}
    \acro{GOP}{Group of Pictures}
    \acro{LL-DASH}{Low-latency DASH}
    \acro{HoL}{head-of-line}
    \acro{IETF}{Internet Engineering Task Force}
    \acro{TSA-SWITCH}{Time Shift-Aware SWITCH}
\end{acronym}

\def\ours{\ac{TSA-SWITCH}}
\def\theirs{SWITCH}

\begin{document}

\title{An Evaluation of ABR Switching for Time-Shifted Clients in MoQ}

\author{
\IEEEauthorblockN{
Abanisenioluwa Orojo\textsuperscript{\orcidlink{0009-0005-9448-1929}},
Tanvir Redoy\textsuperscript{\orcidlink{0009-0000-6078-3463}},
Samira Afzal\textsuperscript{\orcidlink{0000-0003-4779-3936}},
Andrew C. Freeman\textsuperscript{\orcidlink{0000-0002-7927-8245}}
}
\IEEEauthorblockA{
\textit{Department of Computer Science}\\
\textit{Baylor University, USA}
}
}

\maketitle

\begin{abstract}
Media over QUIC enables ultra low latency video streaming over QUIC, but its default quality-switching semantics risk introducing playback gaps during periods of network congestion. The in-progress SWITCH specification for MOQ Transport aims to streamline rate adaptation for MoQ. In this work, we characterize the performance of SWITCH-style Adaptive Bitrate (ABR) for both live and time-shifted clients in a Mininet simulated topology. We validate that standard ABR algorithms can be directly applied to time-shifted playback without modification, yielding substantially higher throughput. We demonstrate that a subscriber can experience increased overall throughput after a rebuffering scenario, and we identify focal points for further optimizations of MoQ ABR switching.

\end{abstract}


\section{Introduction}
Live-streaming events are quickly increasing network burdens, accounting for all of the top ten Internet traffic days in 2024~\cite{applogic2025gipr}. Although \ac{HAS} protocols such as \ac{DASH}~\cite{dash-ISO23009_1_2022} and \ac{HLS} have become ubiquitous for over-the-top video delivery, they have suffered from high latency. Low-latency variants (e.g., LL-DASH, LL-HLS) \cite{dashjs-low-latency-dash, apple-low-latency-hls, bentaleb2022MM} have recently gained prominence for live streaming applications. However, these protocols are encumbered by the pull-based semantics of HTTP, limiting delivery speed and scalability \cite{nguyen2020EPIQ, nguyen2022MHV, nguyen2025MMSys}.

To address these limitations, 
the \ac{IETF} is developing the \ac{MOQT} protocol \cite{moqt-draft} as a flexible and latency configurable framework for diverse applications. \ac{MOQT} uses a publisher/subscriber model with a relay fan-out system, delivering arbitrary temporally-ordered data payloads across WebTransport or raw QUIC. Unlike LL-DASH, \ac{MOQT}'s QUIC-based prioritization schemes ensure that under tight latency budgets higher throughput is achievable \cite{moq-deathmatch}. Hence, \ac{MOQT} is positioned as a better protocol for real-time video streaming applications that demand both low-latency delivery and fine-grained control over stream prioritization.

A user does not always require minimal latency, however. For example, a live broadcast client may experience rebuffering due to network congestion. Rather than jump ahead to the live edge after switching quality, the user may wish to continue playback at the same point, and maintain a delay (time shift) from the live edge. Our own recent work explored a similar scenario with \ac{MoQ}, where a user may choose to have higher playback latency in order to unlock more granular video analysis for content moderation \cite{andrew2025ICME,andrew2026MHV}. 

While single-stream playback is naturally supported in \ac{MOQT} reference software, \ac{ABR} switching semantics are an area of active discussion, research, and development within the \ac{MoQ} Working Group. Although many contributors are focused on live-edge playback, we see that time-shifted playback with \ac{ABR} is understudied and unaddressed in current implementations.   
To better understand the implications of rate adaptation in heterogeneous latency regimes, we introduce an extension to the MOQtail \cite{zafer-moqtail2026mmsys} reference software that enables near-seamless playback with \ac{ABR} for both live and time-shifted clients. The system separates quality selection from timeline synchronization. 

The main contributions of this work are the following:
\begin{itemize}


\item We implement time shift-aware \ac{ABR} switching and multiple \ac{ABR} algorithms in open-source reference \ac{MoQ} software.

\item We produce a Mininet testing environment for reproducible network simulations.

\item We characterize the performance of time shift-aware \ac{ABR} under various network conditions and \ac{ABR} configurations, demonstrating that by adding less than four seconds of stream delay can increase the throughput by four times.


\end{itemize}

Our open-source MOQtail fork and Mininet testing framework are available at \href{https://github.com/BaylorMultimediaLab/moqtail/tree/main}{https://github.com/BaylorMultimediaLab/moqtail/}.

\section{Background and Related Work}
\label{sec:background}

\subsection{Media over QUIC Transport}
\ac{HAS} operates on a pull-based mechanism where clients independently request specific content chunks based on network conditions \cite{timmerer2025has}. Video clients fetch segments, which couple media delivery directly to HTTP request–response cycles, incurring inherent delivery latency. WebRTC delivers real-time media streams over UDP to achieve low latency \cite{carlucci2016gccwebrtc}. However, it is fundamentally engineered for low-latency teleconferencing applications, making it inflexible for broadcast streaming or diverse latency regimes. \ac{MOQT}~\cite{moqt-draft} aims to achieve the best of both worlds: it uses fast, push-based UDP delivery (via QUIC), and it expresses media as named units that relays can cache and fan out in a \ac{CDN}. \ac{MOQT} is an IETF protocol draft specification \cite{moqt-draft} that operates on push-based publish/subscribe semantics. Data units are delivered as \emph{Objects} grouped into \emph{Groups} within named \emph{Tracks}, carried over QUIC streams or datagrams via relays~\cite{moqt-draft}. \ac{MOQT} is associated with two video streaming format specifications: the \ac{MSF}, which defines the catalog (a manifest track defining the publisher's available media representation Tracks) and mandates that Tracks in a common render group be synchronized and switchable at Group boundaries~\cite{msf-draft}; and \ac{CMSF}, which synchronizes \ac{CMAF} fragment boundaries with \ac{MOQT} Group boundaries and carries \ac{CMAF} chunks as \ac{MOQT} Objects~\cite{cmsf-draft}. Under these formats, a video frame maps to an Object, a \ac{GOP} maps to a Group, and a given video representation or quality level maps to a Track. 


\subsection{Switching Mechanics and Timeline Continuity}
\ac{MSF}/\ac{CMSF}~\cite{msf-draft,cmsf-draft} establishes that a Group boundary is the join point at which a subscriber may enter a Track. \ac{MOQT} offers three ways to act on it. The naive approach is to issue a \texttt{SUBSCRIBE} message for the new Track and an \texttt{UNSUBSCRIBE} message for the old one; However, because \texttt{UNSUBSCRIBE} stops future delivery but does not cancel Groups already queued or in flight, the old Track can continue delivering while the new Track begins. The two subscriptions overlap, duplicating Group IDs and producing a bandwidth spike during the switch that competes with the new Track and can stall playback. A downside of this approach is that the new subscription only receives Groups published after the subscription starts, so any media already elapsed on the target Track is not delivered. A Joining FETCH~\cite{moqt-draft} lets a client recover older objects from a Track while a SUBSCRIBE keeps receiving new objects from that same Track. In other words, FETCH can fill in missing past data, and SUBSCRIBE can continue forward delivery. However, in the context of \ac{ABR}, FETCH is entirely client-driven, requiring each client to mitigate potential overlap or gaps between the \texttt{UNSUBSCRIBE} and \texttt{FETCH} messages.

In contrast, the pending MOQ SWITCH ~\cite{gurel-moq-track-switching-01} design introduces an atomic relay-side transition from a current Track to a target Track. With SWITCH, the subscriber asks the relay to move an existing subscription from one Track to another. The request identifies the current subscription, the target Track, and the earliest Group where the switch is allowed to occur. The relay then chooses a valid transition Group at or after that point, usually at a Group boundary shared by both Tracks. If that Group is behind the live edge, the relay first sends the missing target-Track data needed to catch up. Once the target Track subscription has begun, the relay ends the prior Track subscription, so the client changes representations through a single relay-managed transition rather than two independent control messages. 

\subsection{Problem Statement}

A gap remains open in the literature surveyed above: \ac{ABR} for the time-shifted regime in \ac{MoQ} is understudied. Existing \ac{MoQ} \ac{ABR} work is focused on optimizing for a fixed live-edge target. In practice, however, many applications will not demand the lowest possible latency at all times. Therefore, we seek to evaluate the efficacy of relay-side \ac{ABR} switching under heterogeneous latency requirements.

\section{System Overview}
\label{sec:system-design}
\begin{figure*}[t]
    \centering
    \begin{subfigure}{0.73\textwidth}
        \centering
        \includegraphics[width=\linewidth]{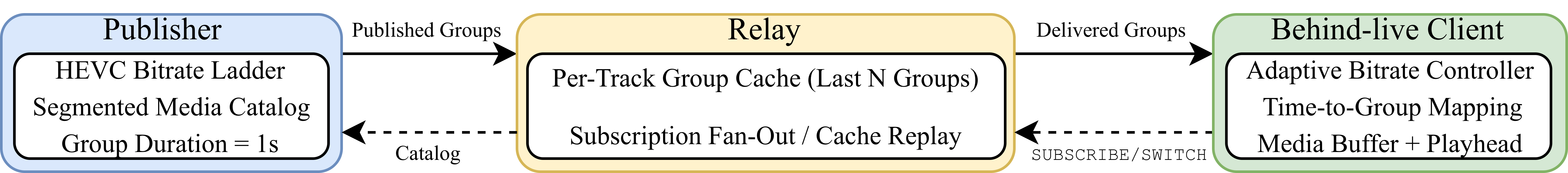}
        \caption{System architecture: publisher, relay cache, and behind-live client.}
        \label{fig:moq-arch}
    \end{subfigure}

    \vspace{0.5em}

    \begin{subfigure}{0.73\textwidth}
        \centering
        \includegraphics[width=\linewidth]{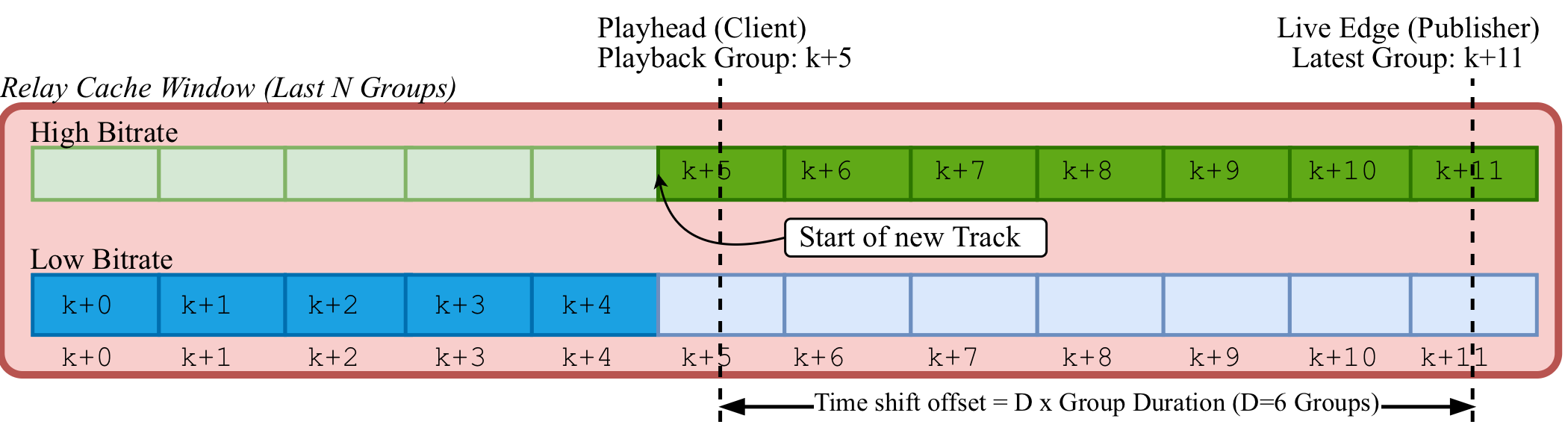}
        \caption{Group timeline example showing playback delay offset, and track switching at the playhead.}
        \label{fig:moq-timeline}
    \end{subfigure}

    \caption{Overview of the MOQ-based adaptive streaming design.}
    \label{fig:moq-overview}
\end{figure*}

\cref{fig:moq-arch} illustrates our system architecture. Our modifications from a typical \ac{MoQ} implementation are primarily to the relay and subscriber, to support controlled playback offsets and \ac{ABR} algorithms. 

\textit{Publisher.} The publisher ingests a source video and constructs a \ac{HEVC} bitrate ladder with a fixed \ac{GOP} duration. Frames and \ac{GOP}s are sent to the relay as \ac{MOQT} Objects and Groups, respectively, according to the \ac{MSF} protocol.

\textit{Relay.} The relay accepts \texttt{SUBSCRIBE} messages from clients and forwards data from the subscribed Tracks. For each Track, the relay maintains a bounded cache of the $N$ most recent Groups. This cache allows the relay to serve both live clients, which receive Groups immediately, and time-shifted clients, which receive Groups from the cache.

\textit{Time-shifted subscriber.} The subscriber client maintains a single active video Track subscription and runs an \ac{ABR} controller that issues representation-change requests in response to throughput, buffer occupancy, and per-frame latency signals. In the time-shifted configuration considered in this work, the client can begin playback from $D$ Groups behind the publisher's live edge. We represent the live edge by $L$, the latest Group currently available at the relay. For example, in \cref{fig:moq-timeline}, the live edge is at Group $L=k+11$, while the client's playhead is at Group $k+5$. The client is therefore $D=6$ Groups behind the live edge. With a 1-second GOP duration, this corresponds to a 6-second playback offset.

\section{TSA-SWITCH}

This section describes our switch implementation, known as \ours{}. We note that the SWITCH mechanism for \ac{MOQT} is still under active discussion in a pull request at the time of writing \cite{simon_switch_2026}, and the current MOQtail implementation of SWITCH only supports \ac{ABR} for live edge clients, so we refer to it as Live-SWITCH. As such, we extend the MOQtail implementation to add simplified support for time-shifted \ac{ABR}. Whereas the current SWITCH \cite{simon_switch_2026} specification draft cedes some authority to the relay to determine the best Group to deliver, our implementation allows only the client to specify the intended Group to receive during a quality switch.

\subsection{Delayed subscription} First, we introduce a \texttt{DELAY\_GROUPS} subscription parameter, included in the client's \texttt{SUBSCRIBE} request to specify a playback delay of $D$ Groups behind the live edge. After receiving the request, the relay computes the target Group as $g_{\mathrm{target}} = L - D$. This target Group determines where delivery should begin. The relay then handles the request in one of three ways. If $g_{\mathrm{target}}$ is available within the relay cache, the relay starts forwarding media from that Group. If $g_{\mathrm{target}}$ is older than the oldest cached Group, the relay starts from the oldest Group still available in the cache. If the stream has not yet produced enough Groups to satisfy the requested delay, the relay holds the subscription pending, activating it only when the live edge advances enough for $g_{\mathrm{target}}$ to become available.



\subsection{Timeline-to-Group mapping} Second, we introduce a \texttt{TimeMap} data structure that the client maintains to record Group boundaries as objects arrive. The \texttt{TimeMap} maintains a sparse mapping from media \ac{PTS} to Group ID. When the \ac{ABR} controller requests a representation switch, the client captures the current playhead \ac{PTS}, obtained from the video element's \texttt{currentTime}. The client then queries the \texttt{TimeMap} for the Group containing that PTS, and attaches the result as the \texttt{START\_LOCATION\_GROUP} parameter in the \texttt{SWITCH} message. In \ours{}, the client sends the playhead-derived Group ID as a \texttt{START\_LOCATION\_GROUP} parameter that approximates the Minimum Switching Group ID from the draft SWITCH design \cite{simon_switch_2026}. The relay uses this value to create an absolute-start subscription for the target Track, and begins sending from that requested Group. In the pending SWITCH draft \cite{simon_switch_2026}, the equivalent signal is carried as a SWITCH field, and the relay may choose a later valid \texttt{G\_switch} if required by common-boundary or cache-availability constraints.



\subsection{Motivation}

We emphasize that our goal at this stage is not to propose a competing SWITCH design. Instead, we aim to provide early experimental results for the \ac{ABR} performance of clients under heterogeneous latency regimes, which may inform future refinements to SWITCH and \ac{ABR} specifications within the \ac{MoQ} Working Group.

\section{Experimental Setup}

\subsection{Implementation}

To evaluate \ac{ABR} for time-shifted playback, we build upon the MOQtail \cite{zafer-moqtail2026mmsys} reference software, which, at the time of our evaluation, tracked Draft 14 of the \ac{MOQT} protocol specification \cite{ietf-moq-transport-14}.

\begin{figure*}[!t]
  \centering
  \begin{subfigure}[t]{0.27\linewidth}
    \centering
    \includegraphics[width=\linewidth]{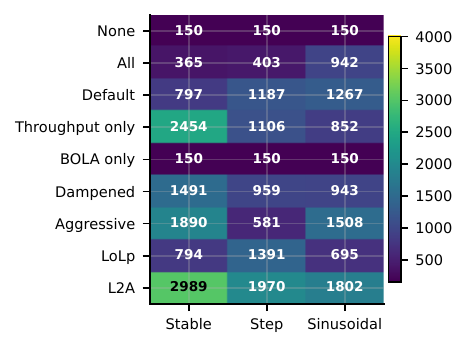}
    \caption{Received bitrate (kbps)}
    \label{fig:bitrate_live}
  \end{subfigure}
  \hfill
  \begin{subfigure}[t]{0.27\linewidth}
    \centering
    \includegraphics[width=\linewidth]{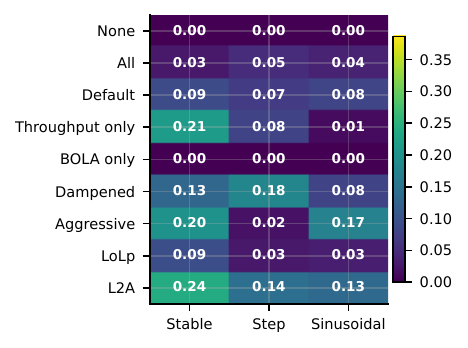}
    \caption{Rebuffering ratio}
    \label{fig:stall_live}
  \end{subfigure}
  \hfill
    \begin{subfigure}[t]{0.27\linewidth}
    \centering
    \includegraphics[width=\linewidth]{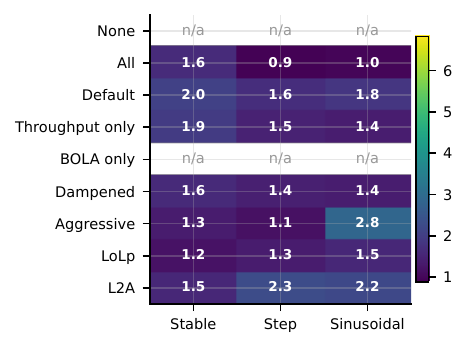}
    \caption{Time to switch (s)}
    \label{fig:time_live}
  \end{subfigure}
  \caption{Average results for Live-SWITCH on the live edge.}
  \label{fig:live}
\end{figure*}

\begin{figure*}[!t]
  \begin{subfigure}[t]{0.27\linewidth}
    \centering
    \includegraphics[width=\linewidth]{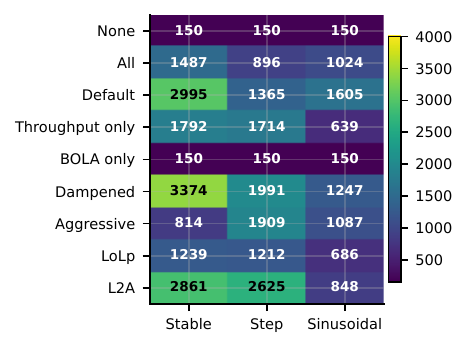}
    \caption{Bitrate requested (kbps)}
    \label{fig:bitrate_shift_live}
  \end{subfigure}
  \hfill
  \begin{subfigure}[t]{0.27\linewidth}
    \centering
    \includegraphics[width=\linewidth]{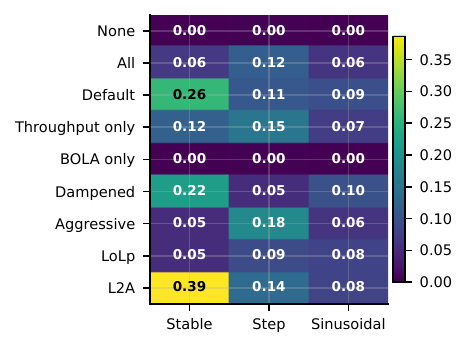}
    \caption{Rebuffering ratio}
    \label{fig:stall_shift_live}
  \end{subfigure}
  \hfill
   \begin{subfigure}[t]{0.27\linewidth}
    \centering
    \includegraphics[width=\linewidth]{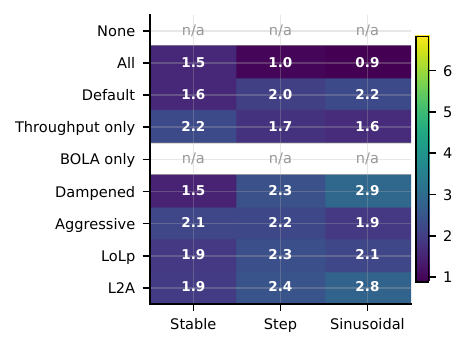}
    \caption{Time to switch (s)}
    \label{fig:time_shift_live}
  \end{subfigure}
  \caption{Average results for \ours{} beginning at the live edge.}
  \label{fig:shift}
\end{figure*}

\begin{figure*}[!t]
  \begin{subfigure}[t]{0.27\linewidth}
    \centering
    \includegraphics[width=\linewidth]{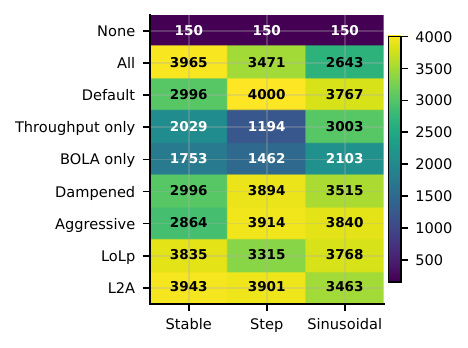}
    \caption{Bitrate requested (kbps)}
    \label{fig:bitrate_shift}
  \end{subfigure}
  \hfill
  \begin{subfigure}[t]{0.27\linewidth}
    \centering
    \includegraphics[width=\linewidth]{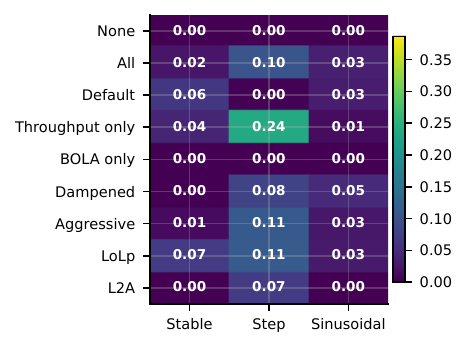}
    \caption{Rebuffering ratio}
    \label{fig:stall_shift}
  \end{subfigure}
  \hfill
   \begin{subfigure}[t]{0.27\linewidth}
    \centering
    \includegraphics[width=\linewidth]{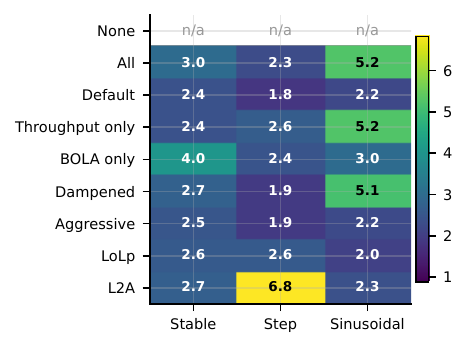}
    \caption{Time to switch (s)}
    \label{fig:time_shift}
  \end{subfigure}
  \caption{Average results for \ours{} with a delayed starting offset of 10 s.}
  \label{fig:shift}
\end{figure*}

\subsection{Video Sequence and Ladder}
To avoid transcoder bottlenecks and ensure reproducibility, we pre-encode the video segments. We take the first 60 seconds of Tears of Steel \cite{tearsofsteel} 
encoded as a five-rung HEVC ladder, with target bitrates of \{150, 200, 500, 1200, 4000\}\, kilobits per second (kbps). The top rung is deliberately set above our tested bandwidth ceiling so that down-switching is induced under the constrained profiles. We set our GOP duration to 1 s.

\begin{table}[!t]
\centering
\scriptsize
\begin{tabularx}{\linewidth}{  l X X  }
\toprule
\textbf{Configuration} & \textbf{Rules Enabled} & \textbf{Override Settings} \\
\midrule
\emph{None} & None & \\\hline

\emph{All} &
All dash.js ABR rules in the player ~\cite{dashjs,dashjs-abr-settings} &
 \\\hline

\emph{Throughput Only} &
ThroughputRule ~\cite{dashjs-throughput-rule} &
 \\\hline

\emph{BOLA Only} &
BolaRule ~\cite{spiteri2016bola,spiteri2019dash} &
 \\\hline

\emph{Default} &
ThroughputRule, BolaRule ~\cite{dashjs-abr-settings,spiteri2019dash} &
 \\\hline

\emph{Dampened} &
ThroughputRule, BolaRule, SwitchHistoryRule~\cite{dashjs-abr-settings,dashjs-switch-history-rule} &
Tightened switch-history threshold \\\hline

\emph{Aggressive} &
ThroughputRule, BolaRule~\cite{dashjs-abr-settings} &
\texttt{bandwidth Safety Factor}$=1.0$ \\\hline

\emph{LoLp} &
LoL+ ~\cite{lim2020lol,bentaleb2022lolplus} &
 \\\hline

\emph{L2A} &
L2A-LL ~\cite{karagkioules2020l2a} &
 \\
\bottomrule
\end{tabularx}
\caption{ABR configurations evaluated. ``All'' enables all dash.js ABR rules available in the player stack as a stress configuration.}
\label{tab:abr-configs}
\end{table}

\subsection{ABR Configurations}\label{sec:abr-configs}

As detailed in \cref{tab:abr-configs}, our evaluation uses eight \ac{ABR} configurations derived from the rule set exposed by the dash.js reference player~\cite{dashjs}, plus a baseline ``None'' configuration with \ac{ABR} disabled. We use the term \emph{driver} for a rule that directly selects the candidate representation, such as the throughput-based rule, BOLA, L2A, or LoL+. We use the term \emph{support rule} for a rule that modifies or constrains that decision, such as switch-history, insufficient-buffer, dropped-frame, or request-abandonment logic. The \emph{Override Settings} column reports only parameters changed relative to the corresponding baseline configuration. The \emph{All} configuration is a configuration in which all dash.js ABR rules available in our player stack are enabled, whereas \emph{default} corresponds to the dash.js dynamic pairing of throughput-based and BOLA-based decision logic. Where applicable, we preserved the default dash.js playback buffer target of 18 s.

\begin{figure*}
    \begin{subfigure}[t]{0.28\linewidth}
    \centering
    \includegraphics[width=\linewidth,trim={0cm 0 3.2cm 0},clip]{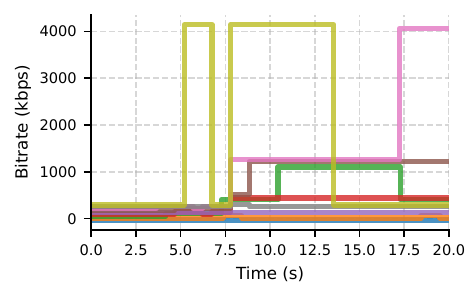}
    \caption{Live-SWITCH on the Live Edge}
    \label{fig:ts_live}
  \end{subfigure}
  \hfill
  \begin{subfigure}[t]{0.28\linewidth}
    \centering
    \includegraphics[width=\linewidth,trim={0cm 0 3.2cm 0},clip]{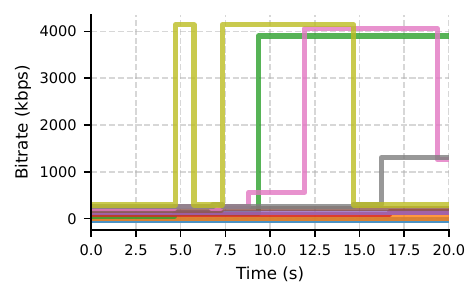}
    \caption{\ours{} beginning on the Live Edge}
    \label{fig:ts_live}
  \end{subfigure}
  \hfill
  \begin{subfigure}[t]{0.38\linewidth}
    \centering
    \includegraphics[width=\linewidth]{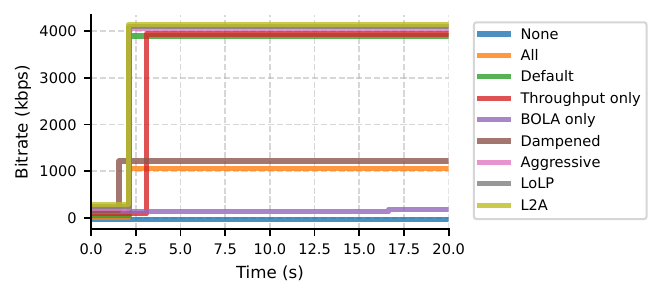}
    \caption{\ours{} with an offset of 10 s}
    \label{fig:ts_live}
  \end{subfigure}
  \caption{Time series plots of representative experimental runs on the Sinusoidal bandwidth profile.}
  \label{fig:ts}
\end{figure*}

\subsection{Testbed} 
All experiments were run under Mininet~\cite{mininet} network simulation. The relay is placed between two emulated switches, and the client runs in a separate network namespace using headless Chromium 148.0.7778.179. Both links are \texttt{TCLink}-shaped with a \qty{10}{\milli\second} one-way delay
; bandwidth on the relay-to-client link ($s_2$) is varied at runtime by the experiment harness. Link 1 (publisher-to-relay) is held at \qty{10}{\mega\bps} and is not the bottleneck for the experiment. To emulate a live source with our pre-encoded bitrate ladder, the publisher paces the sending of these segments to the relay.

\subsection{Bandwidth Profiles}
Three bandwidth profiles were used in the evaluation.
\begin{itemize}
    \item \textit{Stable:} a flat \qty{1.5}{\mega\bps} holding for 
    \qty{60}{\second}. This is sufficient for the four lowest-quality representations.
    \item \textit{Step:} begin at \qty{3}{\mega\bps}, step down to 
    \qty{500}{\kilo\bps} at $t = \qty{30}{\second}$, and recover to 
    \qty{3}{\mega\bps} at $t = \qty{50}{\second}$. This is designed to force a fast 
    down-switch followed by an up-switch.
    \item \textit{Sinusoidal:} a sine wave on $[0.6, 3.0]~\unit{\mega\bps}$ 
    with a \qty{60}{\second} period. This is used to probe ABR behavior under 
    continuously varying capacity without sharp transitions.
\end{itemize}

\subsection{Subscribers}

Our baseline subscriber configuration targets the live edge, performing the Live-SWITCH as implemented in the MOQtail reference software \cite{zafer-moqtail2026mmsys}. We compare to a subscriber with \ours{} beginning at the live edge, and to a subscriber with \ours{} beginning with a delay offset of 10 s.

\subsection{Metrics}

For each experimental run, we measure the following:
\begin{itemize}
    \item Average received video bitrate (kbps).
    \item Ratio of playback time spent rebuffering. This is measured in the client player in 500 ms intervals.
    \item Average time to switch (s). This measures the time elapsed between the client sending a SWITCH message and receiving the first data from the new Track subscription.
\end{itemize}

\section{Results}
\label{sec:evaluation}



\subsection{Live-SWITCH on the Live Edge}
We first examine the scenario of live stream client that wishes to remain as close to the live edge as possible. We configure this by using the Live-SWITCH mechanism as implemented in the MOQtail \cite{zafer-moqtail2026mmsys} reference software. \cref{fig:bitrate_live} shows the average bitrates for our live edge client using the existing MOQtail Live-SWITCH implementation across our experimental test suite. We see quite low throughput overall, and inconsistent \ac{ABR} configuration performance. From \cref{fig:stall_live}, we see that this Live-SWITCH can have substantial rebuffering up to 24\% of the playback duration for \textit{L2A}, due to the small playback buffer. In this configuration, rebuffering causes jumps in the playback to remain at the live edge. The \textit{LoLp} and \textit{All} configurations provide the best all-around performance, yielding high throughput, little rebuffering, and low switch latencies (\cref{fig:live}).

\subsection{\ours{} Beginning on the Live Edge}
We next examine the scenario of a live stream client that is tolerant to increases in latency, through the use of the \ours{} mechanism. In this case, the client begins by targeting the live edge. If rebuffering occurs, the client does not experience substantial gaps in the received video content, and will instead pause and enter a time-shifted state, where the delay is equivalent to the cumulative duration of rebuffers. We see in \cref{fig:bitrate_shift_live} that this yields much higher throughput overall, but the amount of time spent rebuffering (\cref{fig:stall_shift_live}) generally increases. Although the \textit{Dampened} throughput is more than double that of the Live-SWITCH, the rebuffering time is also nearly double. The average rebuffer period for the \textit{All} configuration is less than 4 seconds for the Stable profile, but we see a 4-times improvement in throughput over the Live-SWITCH.

\subsection{\ours{} Beginning with 10-Second Time Shift}
Finally, we explore \ours{} with a 10-second time shift at the outset. As the playhead moves further from the live edge, the client benefits more from the relay's cache and bursty network conditions, yielding consistently higher throughput as in \cref{fig:bitrate_shift}. However, this comes with even greater switch latency (\cref{fig:time_shift}) and in some cases higher rebuffering rates. We see dramatic decreases in rebuffering rates for the Stable and Sinusoidal profiles, but the Step rebuffering remains high, as its dramatic bandwidth change is harder for the \ac{ABR} controller to predict. \cref{fig:ts} illustrates the change in received bitrate over time, showing that the flexibility of a higher latency tolerance allows for quicker rate stabilization and fewer quality switches. 

\subsection{Discussion}
\label{sec:discussion}
Our experimental observations are in line with expectations for \ac{ABR} switching across different latency regimes. We see that the reduced rebuffering at higher delay offsets corresponds to the greater headroom afforded by the relay's cache availability. 
Unsurprisingly, the \textit{All} configuration maintains strong performance across our bandwidth profiles as the offset increases. We find that the overall switch latency is higher than expected, especially for the live-oriented clients. 
Under \ours{}, if switch executes later than the next GOP begins, the client receives redundant (past) data which the player discards. Ideally, the time to switch will be less than or equal to the \ac{GOP} duration, to support faster reactions to changes in network conditions. Since the control streams used for the SWITCH messages are already assigned the highest QUIC stream priority, this latency is likely due to the relay-side overhead of initiating the new Track subscription. This is a target for further software optimization.

\section{Conclusion}
\label{sec:conclusion}

We evaluated early \ac{ABR} semantics within \ac{MoQ}, exploring heterogeneous latency regimes with a single SWITCH implementation (\ours). We demonstrated that a time-shifted client seamlessly provided more adaptation headroom, enabling higher and more consistent delivered quality across the tested network profiles, frequently approaching the maximum bitrate target of 4000 kbps. We identify opportunities to optimize the relay-side SWITCH execution time, and we anticipate the development of advanced hybrid \ac{ABR} systems, where the relay coordinates with the client to improve \ac{ABR} decision-making (e.g., suggesting algorithm parameter adjustments based on its own connection to the upstream publisher). Towards this end, we provide an open-source Mininet integration to streamline further evaluation of \ac{MoQ} \ac{ABR} with repeatable experiments. With this foundation, an extended evaluation can compare SWITCH-based \ac{ABR} to Joining FETCH and relay-side switching. Ongoing research will inform the development of the SWITCH design and its implementation in reference software, helping to enable the adoption of \ac{MOQT} for live streams, video on demand, and everything in between.

\bibliographystyle{ieeetr}
\bibliography{references}

\end{document}